\documentclass[aps,prl,superscriptaddress,showpacs,twocolumn]{revtex4}
\usepackage{amsfonts}
\usepackage{amsmath}
\usepackage{amssymb}
\usepackage{graphicx}

\begin{document}

\title{Effect of next-nearest neighbor coupling on the optical spectra in bilayer graphene}
\author{A. R. Wright}
\affiliation{School of Engineering Physics, University of
Wollongong, New South Wales 2552, Australia}
\author{Feng Liu}
\affiliation{Department of Materials Science and Engineering,
University of Utah, Salt Lake City, UT 84112, USA}
\author{C. Zhang}
\affiliation{School of Engineering Physics, University of
Wollongong, New South Wales 2552, Australia}

\begin{abstract}
We investigate the dependence of the optical conductivity of
bilayer graphene (BLG) on the intra- and inter-layer interactions
using the most complete model to date. We show that the next
nearest-neighbor intralayer coupling introduces new features in
the low-energy spectrum that are highly sensitive to sample
doping, changing significantly the ``universal'' conductance.
Further, its interplay with interlayer couplings leads to an
anisotropy in conductance in the ultraviolet range. We propose
that experimental measurement of the optical conductivity of
intrinsic and doped BLG will provide a good benchmark for the
relative importance of intra- and inter-layer couplings at
different doping levels.

\end{abstract}
 \maketitle

Since the isolation of single layers of graphite in 2003
\cite{novo1}, a lot of exciting work on single layer graphene
(SLG) has been done \cite{novorev}. For example, the prediction
and observation of electron-hole symmetry and a half-integer
quantum Hall effect \cite{novo2,zhang1,berg},  finite conductivity
at zero charge-carrier concentration \cite{novo2}, the strong
suppression of weak localization
\cite{suzuura,morozov,khveshchenko}, universal conductance
\cite{gus,kuz,nair} and magnetic enhancement of optical
conductance in graphene nanoribbons \cite{liu1}.

More recently, attention has also been paid to SLG's cousin,
bilayer graphene (BLG). The electronic and transport properties of
BLG differ significantly from SLG in many respects, particularly
at low energies in the `Dirac' regime. Various models for low
energy BLG exist in the literature depending on the coupling terms
included, and whether electronic bands beyond the lowest energy
subbands are retained \cite{nilscast, mccann_prl}. Many
interesting results were obtained based on a model that includes
only the most dominant of the interlayer coupling terms in BLG, as
well as the usual nearest  neighbor intralayer term
\cite{castro1}. By including the second most dominant interlayer
coupling, some unusual properties such as a peculiar Landau-level
spectrum have been derived \cite{mccann_prl}, as well as a new low
energy peak \cite{abe}.

The `universal conductance' of graphene is both a DC and an AC
phenomenon. It is a direct result of the linear energy dispersion
of graphene. Linear subbands imply both a constant density of
states as well as consistent transition matrix elements, which
means that for as long as the linear (Dirac) approximation is
valid, the conductance is a constant. In the AC case, the value of
the universal conductance of single layer graphene is $\sigma_1 =
e^2/4\hbar$. In the layered case, a standard benchmark is simply
$\sigma_n = n\sigma_1$. However, this is not generally accurate,
as the subband curvature caused by interlayer coupling in the case
of layered graphene leads to a non-constant conductivity. This
raises an important question: in what energy range is $\sigma_n =
n\sigma_1$ applicable?

The infrared conductance of BLG has been measured by several
groups \cite{zqli, mak}. These results rely upon the effects of an
induced gate voltage on the bandstructure, and all assume a
discrepancy in onsite energy between the two layers. In ref. 17,
Mak et al present the `expected' IR conductance without the latter
assumption, and find that it differs markedly from their
experimental results. This demonstrates the need to assume an
energetic discrepancy between the two layers in BLG. Our
theoretical results, however, show a strong correlation to the
results in ref.16 and 17, demonstrating that while an energetic
discrepancy may exist, it is not necessary in describing the IR
response observed experimentally.

In this letter, we study the dependence of the optical conductance
of BLG on various intra- and inter-layer couplings. It is shown
that the interplay of these couplings leads to a significant
deviation in the behaviour of the conductance at low frequencies,
which can, in turn, be tuned by electronic doping. In the
important ultraviolet frequency band, this interplay leads to
significant conductance anisotropy, i.e., the absorption along the
armchair direction is around 50\% stronger than that along the
zigzag direction.

A typical BLG sheet consists of two SLG layers stacked in the
orientation shown in Fig.1. Several forms of the Hamiltonian
for BLG are used in the literature depending on the approximations
used and the relative orientations of the two layers. The original
consideration was given by Slonczewski-Weiss-McClure which
included all three interlayer coupling terms \cite{mcclure,sl-we}.
The most prominent interlayer term is the A-B and B-A coupling between
sites which are directly above (or below) each other. Here we
define this term as $\gamma_1 = 0.36eV$. The other two interlayer
coupling terms are the A-B and B-A coupling between inequivalent sites which are not
directly above or below each other, but offset by an amount $b =
1.42\dot{A}$, and the A-A and B-B terms which are similarly offset from
one another, but represent equivalent sites in the SLG Brillouin
Zone. These coupling terms are defined here as $\gamma_3 =
0.10eV$ and $\gamma_4 = 0.12eV$, respectively. We have also
included the next nearest neighbor A-A and B-B coupling which we
define as $t' = 0.30eV$. Finally, as usual, the nearest neighbor
A-B and B-A coupling is included, which is given here by $t =
3.0eV$. All energies will be normalised relative to the first
nearest neighbor coupling.

\begin{figure}[tbp]
    \centering\includegraphics[width=8cm]{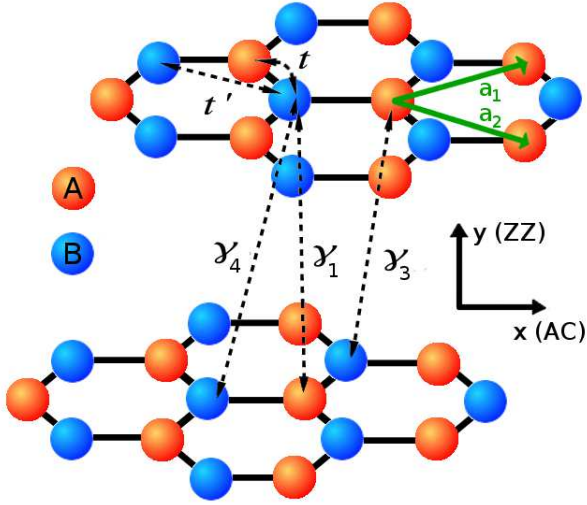}
    \caption{The three interlayer and two intralayer coupling terms included in the BLG Hamiltonian. $\gamma_3$
and $\gamma_4$ differ in that they connect, respectively,
inequivalent (eg. A-B) and equivalent (eg. A-A) points in the SLG
Brillouin Zone. $\gamma_1$ is a directly vertical transition, and
so the overlap of the wavefunctions is about $3\times$ larger than
$\gamma_3$  and $\gamma_4$. The armchair (AC) direction is given
by the x-axis, and the zig-zag (ZZ) direction is given by the
y-axis. The lattice vectors $\mathbf{a}_1$ and $\mathbf{a}_2$ are
also shown.}
\end{figure}

The full Hamiltonian matrix for the BLG system is

\begin{equation}
H_{\mathrm{BLG}} =
\begin{pmatrix}
t'H' & tH^* & \gamma_4 H & \gamma_1\\
tH & t'H' & \gamma_3 H^* & \gamma_4 H \\
\gamma_4 H^* & \gamma_3 H & t'H' & tH^*\\
\gamma_1 & \gamma_4 H^* & tH & t'H'
\end{pmatrix},
\end{equation}
where $H = e^{ik_ya/\sqrt{3}}(1 + e^{i\mathbf{k}\cdot\mathbf{a}_+}
+ e^{i\mathbf{k}\cdot\mathbf{a}_-}) $ and $ H' =
2(\cos(\mathbf{k}\cdot\mathbf{a}_+) +
\cos(\mathbf{k}\cdot\mathbf{a}_-) +
\cos(\mathbf{k}\cdot(\mathbf{a}_+ - \mathbf{a}_-)))$. Here
$\mathbf{a}_\pm = a\bigl(\pm\frac{1}{2},-\frac{\sqrt{3}}{2}\bigr)$
are the two lattice vectors shown in Fig.1. The eigenvalues and eigenvectors in
the absence of $\gamma_4$ are readily solved. With $\gamma_4$ included
however, the form of the solution is unwieldy. The eigenvalues in
the simpler case are given by the (relatively) concise form

\begin{equation}
\begin{split}
\epsilon_{s,s'} &= t'(\epsilon_{\mathrm{SLG}}^2 - 3) +
s \sqrt{ \epsilon_{SL}^2 + \frac{\gamma_{12}^+}{2} + s' \sqrt{\Gamma} }
\end{split}
\end{equation}
Where

\begin{equation}
\Gamma = \epsilon_{SL}^2\gamma_{12}^+ + \frac{(\gamma_{12}^-)^2}{4} + 2\gamma_1\gamma_3\epsilon_{SL}^2\mathrm{Re}(H)
\end{equation}

And $\gamma_{12}^\pm = \gamma_3^2\epsilon_{SL}^2 \pm \gamma_1^2$, with $s,s' = \pm1$, and $\epsilon_{\mathrm{SL}}$
are the regular eigenvalues for the SLG system given as
\begin{equation}
\begin{split}
\epsilon_{\mathrm{SLG}} = t&\bigl(1 +
4\cos(ak_x/2)\cos(ak_y/2\sqrt{3})\\ & +
4\cos^2(ak_x/3)\bigr)^{\frac{1}{2}},
\end{split}
\end{equation}
From this result we see that there are two conduction bands and
two valence bands which are confined above and below the line
$\epsilon_{s,s'} - t'(\epsilon_{\mathrm{SLG}}^2 - 3)$,
respectively. This simple result will form the basis for much of the discussion to follow.

The electron field operators can be constructed from the
eigenvectors $\psi_{s,s'}(\mathbf{k})$ such that
$\Psi(\mathbf{r})
=(1/4\pi^2)\sum_{\mathbf{k},s,s'}a_{ s,s'}(\mathbf{k})
\psi_{ s,s'}(\mathbf{k})e^{i\mathbf{k}\cdot\mathbf{r}}$,
where $a_{ s,s'} (\mathbf{k})$
$(a^\dagger_{ s,s'}(\mathbf{k}))$ denotes the
annihilation (creation) operator for an electron in the $s$ or $s'$
subband with momentum $\mathbf{k}$.

The band structure of BLG near the K points varies dramatically
depending on the coupling terms included in the Hamiltonian. The
effect of the various coupling terms are as follows:

\begin{figure}[tbp]
    \centering\includegraphics[width=8cm]{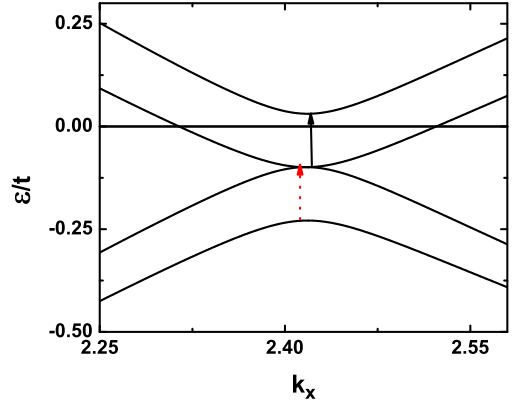}
    \caption{The $k_x$ dependence of the bandstructure near the K/K'
points. The red dashed arrow represents a transition which is
permitted in an undoped sample if NNN coupling is neglected, but
becomes forbidden when it is. The black solid arrow is the
opposite: a previously forbidden transition becomes allowed when
NNNs are included. The effect of  doping is to raise or lower the
Fermi level, making the inclusion of NNNs partly equivalent to
doping.}
\end{figure}

The next nearest neighbor term $t'$ defines the line about which
the conduction and valence bands are confined, as can be seen in
Fig.2. In the absence of lateral warping, the valence and
conduction bands are symmetric about this line. Near the K points,
the prefactor of the energy term associated with $t'$ approaches
$3$, and so the effect of this term is to shift all four bands
downward by an amount $3t'\approx 0.1\mathrm{eV}$. This can easily
be seen from equation 2,  where $H' \approx 3$. This breaks the
electron-hole symmetry, but not the location of the minima in
momentum space.

The term $\gamma_1$ represents the dominant interlayer $A-B$ and $B-A$ coupling.
This term causes an energy gap to form between the two conduction
bands, and an identical gap between the two valence bands of $\gamma_1 \approx 0.13eV$. $\gamma_1$ also removes the
linear dispersion at low energies. The electron hole symmetry is
retained, and no lateral warping occurs. The effect of $\gamma_1$ is apparent in Fig.2.

The second interlayer coupling term $\gamma_3$ restores the linear
lowest energy subband. This term causes what is usually
referred to as `trigonal warping' \cite{mccann_prl, trig_effect}. A second set of Dirac points
near the K/K' points emerges with $\gamma_3$ included, as can be seen in Fig.3.

\begin{figure}[tbp]
    \centering\includegraphics[width=8cm]{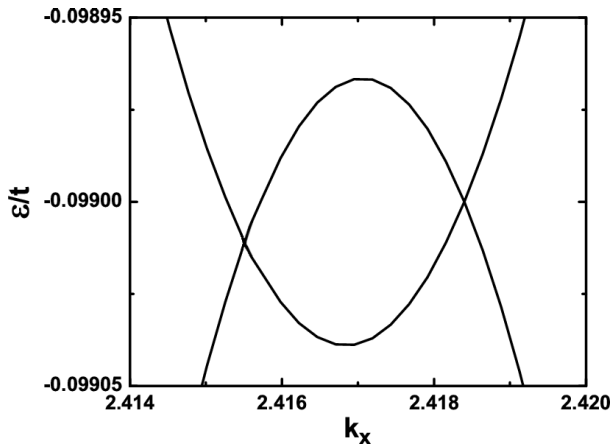}
    \caption{The $k_x$ dependence of the two inner bands near the K/K'
points zoomed right in to see the effects of the intralayer next
nearest neighbors (NNN), and interlayer coupling terms $\gamma_3$
and $\gamma_4$. The NNN interaction has shifted these features
well below the Fermi level. $\gamma_3$ causes a second dirac point
to emerge, and $\gamma_4$ skews the bandstructure, causing one of
the two Dirac points to be pushed down to a lower energy.}
\end{figure}

Finally, $\gamma_4$ causes one of the Dirac points to be plunged
below the NNN line, also seen in Fig.3. When the next nearest
neighbor  term couples with $\gamma_4$, however, the low energy
x-y isotropy is substantially weakened. While $\gamma_3$ causes
the well known anisotropic `trigonal warping', the energy range of
this effect is in the order of $t/10000\approx 0.0003eV$. On its
own, the effect of $\gamma_4$ is similarly small. Here, however,
$\gamma_4$ and $t'$ both couple equivalent sites, which causes a
compounding of their individual effects on the electronic
dispersion relation. The effect is quite a large deviation from
isotropy. The effect of this deviation is most noticeable in the
low energy conductance anisotropy shown in Fig.5.

We now evaluate the optical conductivity of BLG in the absence of
disorder or impurities, over all relevant photon energies. By
using the Kubo formula, the optical conductivity is given as \cite{mahan},

\begin{equation}
\sigma_{\mu,\nu}(\omega) =
\frac{1}{\omega}\int_0^\infty\mathrm{d}te^{i\omega t}
\langle[J_\mu(t),J_{\nu}(0)]\rangle.
\end{equation}

The components of the current operator can be calculated from
$J_{\nu,\mu}(t) = e^{iHt} J_{\nu,\mu}(0) e^{-iHt}$,  where
$J_{\nu,\mu}(0) =
\Psi^\dagger(\mathbf{r})\widehat{v}_{\nu,\mu}\Psi(\mathbf{r'})$,
in which $\widehat{v}_{\nu,\mu} = \partial H/\partial
k_{\nu,\mu}$, and ${\nu,\mu} = x,y$. These values are calculated
numerically, but we note that for each band there are three types
of interband transitions, and also intraband transitions. In the
case of no disorders and no intermediate interactions, it is found
that intraband transitions cannot occur.

In Fig.4. we examine the full energy optical conductance of
undoped bilayer  graphene with NNN interactions included. Near the
higher energy valley points, the optical conductivities exhibit
two extrema, similar to the single peak found in single layer
graphene \cite{chao-ma, stauber}. These peaks correspond to the
two dominant vertical transitions between the two symmetric pairs
of saddle points. The Joint Density of States in these valleys
reaches a cusp-like maximum which leads to the extrema in the
conductivities. These two energy peaks are separated by an amount
$\hbar\omega = 2\gamma_1$, as expected from the bandstructure
calculations.

\begin{figure}[tbp]
    \centering\includegraphics[width=8cm]{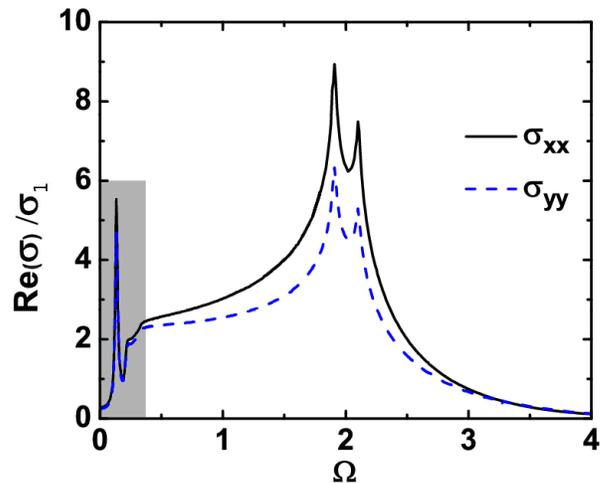}
    \caption{The full energy optical conductance (in units of $\sigma_1=e^2/\hbar$)
vs the normalized frequency $\Omega=\hbar\omega/t$ for bilayer
graphene. Generally, $\sigma_{xx}$ (the armchair direction) has a larger optical response
than $\sigma_{yy}$ (the zig-zag direction). When NNN and $\gamma_4$ are neglected, and at low energies,
$\sigma_{xx} = \sigma_{yy}$. This is no longer the case here, with NNNs and $\gamma_4$
included. For $\epsilon < 3t'$, there are no allowed transitions
and the OC is approximately zero. The grey shaded area indicates the low energy region plotted in Fig.5.}
\end{figure}

\begin{figure}[tbp]
    \centering\includegraphics[width=8cm]{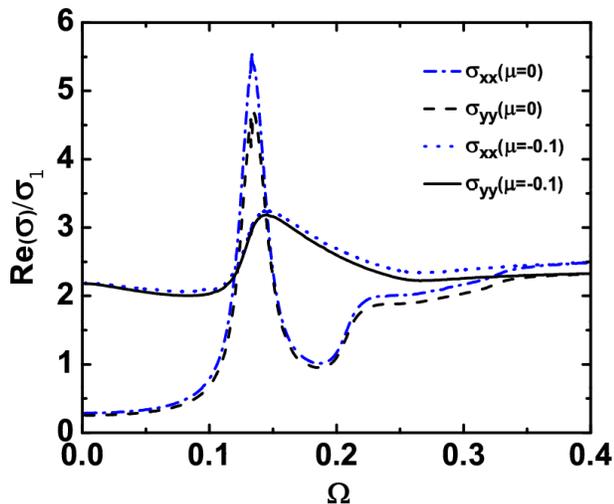}
    \caption{The low energy optical conductance region from Fig.4.
    at two different doping levels. The black solid line is the OC of undoped BLG.
    The blue dotted line represents the OC at finite doping, and is equivalent to
    an undoped sample with no NNN interaction included which has been previously reported. The NNN-$\gamma_4$
coupling causes a new peak to emerge, and suppresses the
previously reported one. This new peak is much larger and shifted to a lower photon energy. In a
suitably doped sample, however, the $t'=0$ (no NNN) peak has been retrieved
by an effective shifting of the Fermi level.}
\end{figure}

Fig.5. shows the low energy optical conductivity of the undoped
sample from the grey shaded region of Fig.4., as well as a sample
doped to the level of the NNN interaction. The longitudinal
conductivities vary greatly when including a non-zero NNN
interaction. The effect of the dominant interlayer term $\gamma_1$
(ie. setting $\gamma_3=\gamma_4=t'=0$) at low energies has been
reported recently \cite{abe}. This result has been retrieved in
our result by doping the sample to $\mu = -0.1eV/t$. In the
undoped curve however, the previous result is entirely suppressed,
and replaced by an approximately $2\times$ larger, rounded peak,
followed by a significant trough. This correlates well with the
behaviour observed in IR experiments \cite{zqli,mak}, although
without the added effects of an induced gate voltage. The doped
case represents a transition of electrons from the upper
conduction band to both the upper and lower valence bands, which
is equivalent to neglecting the NNN coupling, as shown in Fig.2.
The undoped case however reflects the suppression of transitions
into the lower valence band, since it is already filled, and yet a
new set of transitions occur between the two valence bands. These
bands are separated by an approximately constant factor of
$\gamma_1$, which leads to the large peak centred at $\hbar\omega
\approx \gamma_1$ in the undoped bilayer case. The feature is in
striking contrast with that of SLG. For SLG, the effect of the NNN
coupling is to suppress the universal conductance at low
frequencies \cite{stauber}. As is clearly seen in Fig.5., for BLG
the interplay of the interlayer coupling and the NNN coupling can
suppress the conductance at low frequencies. However, it also
induces a strong absorption peak in the far infrared before the
onset of the universal conductance.

For this reason, the low energy approximations of the behaviour of
bilayer graphene are generally more relevant to doped samples,
with the undoped bilayer properties being drastically affected by
the next  nearest neighbor hopping and additional interlayer
terms. Furthermore, when using existing theories to explain
experimental results, it needs to be noted that an energetic
discrepancy between layers, as well as the inclusion of a gate
voltage, both cause some similar effects to the inclusion of the
NNN interaction. All of these will therefore need to be accounted
for when explaining any experimental result.

Furthermore, the conductance anisotropy observed in Figures 4 and
5,  which is prominent even in the IR region when $\gamma_4$ and
$t'$ are both included, makes the polarization of the photon beam
in experiments a relevant parameter. This orientation dependence
of the optical conductance makes determination of the orientation
of a BLG flake possible, and also makes BLG a potential partial
polarizer. The doping dependence of the low energy conductance
anisotropy makes this feature quite versatile.

Finally, as we have already mentioned, the value of the `universal' conductivity is a topic of great
interest at the moment. According to these results, which have
been  calculated from the most robust interlayer and intralayer
model adopted to date, the value of the universal conductivity is
$\sigma = 2\sigma_1$ where $\sigma_1 = e^2/4\hbar$ is the
universal optical conductivity of single layer graphene defined earlier. The range
over which this value is applicable is greatly affected by the
presence of the NNN interaction, and the electronic doping. In
particular the NNN interaction causes the very low energy optical
response to become negligible, and around the observed peak, the
optical conductance is $\sigma_{\mathrm{peak}} \approx 4\sigma_1$.

In conclusion, we have studied the longitudinal optical
conductivity of BLG with the inclusion of all relevant interlayer
coupling terms and next nearest neighbor intralayer interactions.
The optical conductivity exhibits double peak resonance separated
by an amount $2\gamma_1$ and centered around $\hbar\omega=2t$.  At
low energies, the NNN interaction leads to entirely new behaviour
of the optical conductivity. The results obtained without NNN
coupling, however, can be retrieved by appropriate electronic
doping. The interplay of the NNN-$\gamma_4$ couplings were found
to lead to significant low energy conductance anisotropy which is
strongly doping dependent. Finally, the value of the universal
conductivity with the most robust formalism used to date has also
been determined, and is given by $\sigma_2 = 2\sigma_1$. These
results will be crucial to the experimental testing of accepted
theories on bilayer graphene, and will be useful for potential low
energy electronic and photonic applications of bilayer graphene.

\begin{acknowledgments}
This work is supported by the Australian Research Council.
\end{acknowledgments}

\end{document}